\documentclass[a4paper,11pt]{article}
\usepackage{pos}
\usepackage{caption}
\usepackage{subcaption}
\usepackage{hyperref}
\usepackage{tabularx}
\usepackage{booktabs}

\title{Analysis optimisation for more than 10\,TeV gamma-ray astronomy with IACTs}
 \ShortTitle{Analysis optimisation for energies above 10\,TeV }

\author*[a,b]{Iryna Lypova}
\author[b]{David Berge}
\author[b]{Stefan Klepser}
\author[b]{Dmitriy Kostunin}
\author[b]{Stefan Ohm}
\author[a]{Stefan Wagner}

\affiliation[a]{Landessternwarte, Universit\"{a}t Heidelberg, K\"{o}nigstuhl, 69117 Heidelberg, Germany}

\affiliation[b]{Deutsches Elektronen-Synchrotron (DESY), Platanenallee 6, 15738 Zeuthen, Germany}

\emailAdd{ilypova@lsw.uni-heidelberg.de}

\abstract{The High Energy Stereoscopic System (H.E.S.S.) is one of the currently operating Imaging Atmospheric Cherenkov Telescopes. H.E.S.S. operates in the broad energy range from a few tens of GeV to more than 50\,TeV reaching its best sensitivity around 1\,TeV. In this contribution, we present an analysis technique, which is optimised for the detection at the highest energies accessible to H.E.S.S. and aimed to improve the sensitivity above 10\,TeV. It includes the employment of improved event direction reconstruction and gamma-hadron separation. For the first time, also extensive air showers with event offsets up to 4.5$^{\circ}$ from the camera centre are considered in the analysis, thereby increasing the effective Field-of-View of H.E.S.S. from 5$^{\circ}$ to 9$^{\circ}$. Key performance parameters of the new high-energy analysis are presented and its applicability demonstrated for representative hard-spectrum sources in the Milky Way.}

\FullConference{37$^{\rm{th}}$ International Cosmic Ray Conference (ICRC 2021)\\
		July 12th -- 23rd, 2021\\
		Online -- Berlin, Germany}

\begin{document}
\maketitle

\section{Introduction}
Analysis methods used in the current generation of Imaging Atmospheric Cherenkov Telescopes (IACTs) are typically optimised to have the best performance in the core energy range where most of Very-High-Energy (VHE) gamma-ray sources are bright, i.e. around 1\,TeV. However, there are many important topics -- such as the search for Galactic PeVatrons, the study of gamma-ray production scenarios for sources (hadronic vs. leptonic), extragalactic background light (EBL) absorption studies -- which require good sensitivity at energies above 10\,TeV. At these high energies, the instrument sensitivity is greatly limited by low gamma-ray statistics due to steeply falling source spectra. Thus, it is very important to have an analysis optimised in this challenging energy range.

In the typical event analysis chain, there are several steps, e.g. image preparation and event selection, event reconstruction, background rejection, whose improvement and optimisation can lead to an increase in gamma-ray statistics and as a result to better sensitivity at high energies. One of the selection parameters in the H.E.S.S. analysis chain makes the event selection based on their distance to the camera centre (offset angle) since reconstruction accuracy degrades as event offset increases. At low energies, this cut discards a small fraction of detected showers. However, the effect increases with energy, and above 10\,TeV, nearly 50\% of the detected events are beyond the typical offset threshold. This work investigates the feasibility to use events beyond the described limitation as a way to increase gamma-ray statistics in the high energy range. 

Thus, Section \ref{sec:sec2} of this proceeding introduces the main idea of the high-energy optimised analysis, which benefits from employing  large-offset events. The section also describes challenges of such an approach and the possible way to solve them. Section \ref{sec:sec3} presents the performance obtained for the developed analysis method. 

H.E.S.S. is an array of five Cherenkov Telescopes, consisting of four smaller size telescopes (CT\,1\,--\,4) located in the corners of a square and one large-size telescope (CT\,5) situated in the middle of the array. Since CT\,5 is focused more on low-energy part of the VHE spectrum and has a smaller angular size of the camera, it is not included in the studies presented in this contribution.

\section{Large-offset-event analysis}
\label{sec:sec2}

\begin{figure}[t]
  \centering
  \begin{subfigure}[b]{0.42\linewidth}
    \includegraphics[trim=250 50 200 165, clip, width=1.0\linewidth]{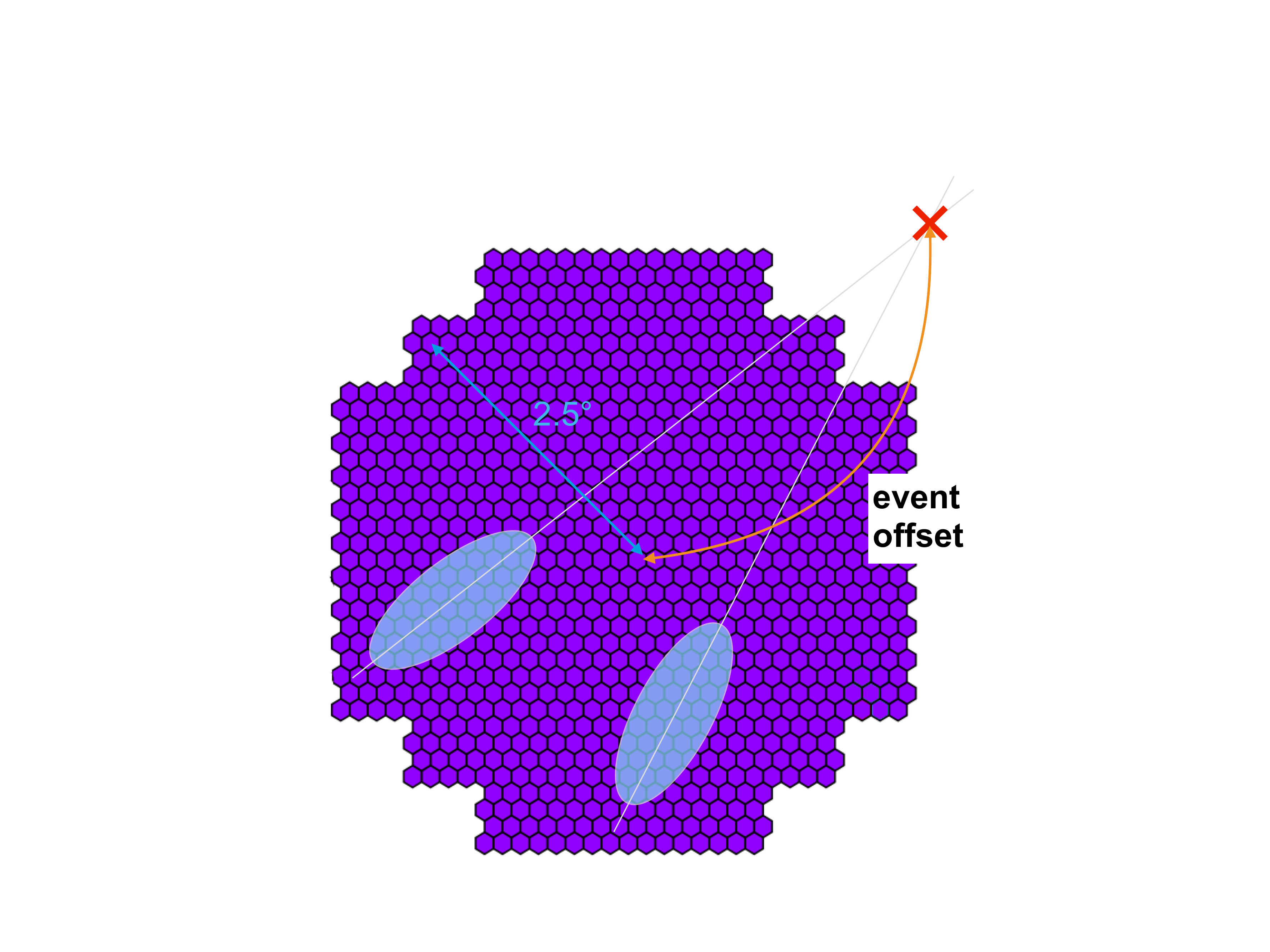}
  \end{subfigure}
    \begin{subfigure}[b]{0.57\linewidth}
    \includegraphics[trim=170 150 200 170, clip, width=1.0\linewidth]{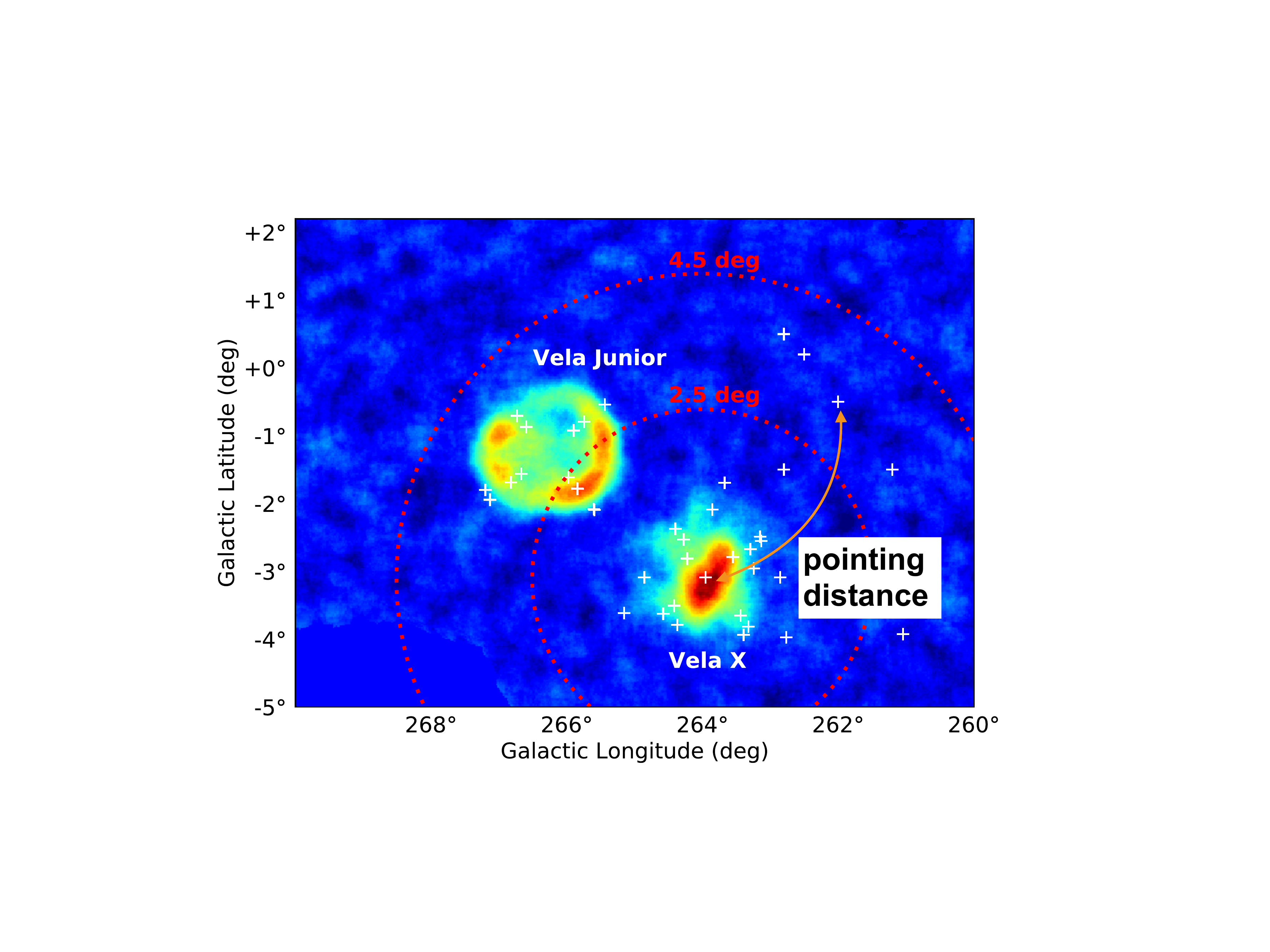}
  \end{subfigure}

  \caption{Left: the camera plane and event offset angle definition. The red cross marks the shower direction reconstructed using the standard method, employing image axes intersection. Right: the definition of the run pointing distance and an example of the sky region where an ability to analyse events at large offset angles would be beneficial. The example displays two VHE sources, Vela X and Vela Junior. White crosses indicate telescope pointing positions during the H.E.S.S. observations. Red dashed rings show two cases of the maximum pointing distance from the Vela X position, i.e. runs within the ring are considered for the analysis. The data are taken from \cite{hgps}.
}
  \label{fig:fig1}
\end{figure}

As mentioned above, event offset is an angular distance between the event direction and camera centre in the camera plane. It is schematically illustrated on the left side of Figure \ref{fig:fig1}. The offset angle is one of the event selection parameters and its standard cut value is normally set to 2.5$^{\circ}$, which is the angular size of the CT\,1\,--\,4 cameras. There is another somewhat related parameter -- pointing distance -- an angular distance between the array pointing position and gamma-ray source that is shown on the right side of Figure \ref{fig:fig1}. It is used for the run selection and the cut value is approximately the same as the cut on the event offset. This way, a simultaneous increase of the maximum allowed event offset and pointing distance can increase the number of observation runs involved in the analysis of the particular VHE source, and thus, increase the source exposure, which may result in larger gamma-ray statistics at high energies. Such an approach is easily realised in the Galactic Plane, where gamma-ray sources are located at relatively small angular distances from each other, and thus, the IACT observation placed rather frequently. The plot on the right in Figure \ref{fig:fig1} shows an example of such a case using Vela X and Vela Junior gamma-ray  sources, which are located within about 3$^{\circ}$ of each other. The smaller ring shows the typical maximum pointing distance of 2.5$^{\circ}$ around the source. About 94\,h of observations fall into this ring. However, for high-energy studies, the pointing distance cut could be increased up to about 4.5$^{\circ}$, which allows for 124\,h of data to be analysed (shown with a larger ring). This way, more runs taken on the neighbouring source (Vela Junior) can contribute to the analysis of the given source (Vela X). In the addition, increase of the maximum allowed event offset results in the increase of the effective Field of View (FoV) from 5$^{\circ}$ to up to 9$^{\circ}$.
 
 
Employment of large-offset events in the analysis has advantages but also comes with a number of challenges. The standard event direction reconstruction \cite{crab} is based on a pairwise intersection of the major image axes and works better for the image pairs with the separation angles between images close to 90$^{\circ}$, which happens fairly often in the case of small-offset events. However, the behaviour of shower and image parameters changes as the event offset increases. Being inclined in relation to the telescope pointing axes, large-offset showers are detectable from greater distances than the small-offset ones. This results in smaller separation angles between shower images. In this case, a small error in direction of the image major axis can lead to the large inaccuracy of the shower direction estimation and degradation of the event analysis accuracy. In addition to this major issue, the efficiency of the gamma-hadron separation drops at large offset angles resulting in an increased background rate being another challenge for the large-offset-event analysis. Both these issues are addressed in the presented work by implementing a more sophisticated direction reconstruction technique within the H.E.S.S. analysis framework as well as modifications to the standard gamma-hadron separation method, which by themselves are the crucial components for the increase of high-energy gamma-ray statistics and improvement of sensitivity.


\subsection{Improved direction reconstruction}
\label{subsec:sec2_1}

Originally, the DISP method (short for displacement) was invented for the shower direction reconstruction in the case of a single telescope operation such as Whipple \cite{disp_whipple}. Nowadays it is widely used for stereoscopic systems such as MAGIC II \cite{disp_magic2}. In the H.E.S.S. analysis framework, the DISP method is implemented for the CT\,5 mono analysis \cite{disp_ct5}. In this work, it is implemented for the CT\,1\,--\,4 telescopes to improve the quality of the shower direction reconstruction. 

The \textit{displacement} is the angular distance between the image centre of gravity (CoG) and the shower direction position. It can be estimated based on the shape of the shower image in the camera. If the vertical shower landed near the telescope, it has a fairly round footprint, while images of distant vertical showers are more elongated. The estimation of the displacement parameter here is similar to the CT\,5 mono reconstruction and done using the machine learning algorithm, specifically Multi-Layer Perceptron (MLP), one of the neural networks implementations within the TMVA package~\cite{tmva}. Seven parameters are used as input variables for the training. Six of them are image parameters: width and length of the image, roundness (width over length), intensity, skewness and kurtosis. The last variable is the seeding parameter. Each event undergoes direction pre-reconstruction using the standard image axes intersection method. The resulting preliminary direction position is used to calculate intermediate displacement value, which serve as a seed for the DISP method.

The DISP method is trained in bands of zenith, azimuth and offset angles as well as optical efficiency. The details of the bins used for the training are summarised in Table \ref{tab:disp_binning}. The neural networks are trained using diffuse gamma rays simulated within the 5$^{\circ}$ viewcone. For the training  in offset-angle bands, this viewcone is split into nine rings with 0.5$^{\circ}$ width (except for the first and last ring, which covers 0$^{\circ}$  to 0.75$^{\circ}$  and 4.5$^{\circ}$  to 5$^{\circ}$ offset angles, respectively). The simulated energy range depends on the zenith angle and spans from around 100\,GeV up to 150\,TeV. To have decent statistics at high energies, gamma rays are simulated with $\Gamma = -1$ spectral index instead of $\Gamma = -2$.


\begin{table}[h!]
\centering
 \begin{tabular}{c | c} 
 \toprule
  \quad Zenith 			\quad & \quad 0, 20, 30, 40, 45, 50, 55, 60$^{\circ}$ \quad \\  
  \quad Azimuth 		\quad & \quad180$^{\circ}$ \quad \\  
  \quad Offset 			\quad & \quad 0.5\,--\,4.5$^{\circ}$ with 0.5$^{\circ}$ step \quad \\ 
  \quad Optical efficiency 	\quad & \quad 50\,--\,100\% with 5\% step \quad \\ 
 \bottomrule
\end{tabular}
\caption{The binning used for the neural network training in the DISP method.} 
 \label{tab:disp_binning}
\end{table}

During the event analysis, the displacement is estimated for each shower image individually. However, the location of the shower direction along the major image axis with respect to the image CoG (in front or behind) is unknown. Thus, each image has two estimated direction positions, whose distance to the image CoG is equal to the displacement estimate. Since CT\,1\,--\,4 telescopes typically observe in stereoscopic mode, this ambiguity is solved by considering all possible combinations of the individual image directions (averaged direction estimates) and picking the one with the smallest uncertainty. Improvement in the angular resolution using the DISP direction reconstruction over the standard method reaches 5\,--\,10\% for the lowest offset-angle band. The effect rises as offset and zenith angle increases. At 2$^{\circ}$ event offset, the improvement is around 20\% and 30\% for 20$^{\circ}$ and 40$^{\circ}$ zenith angle, respectively. Naturally, it happens due to a larger fraction of distant showers with nearly parallel major image axes at large zenith-offset angles than at lower ones.


\subsection{Improved gamma-hadron separation}
\label{subsec:sec2_2}

\begin{figure}[b]
  \centering
  \begin{subfigure}[b]{0.48\linewidth}
    \includegraphics[trim=0 0 0 18, clip, width=1.0\linewidth]{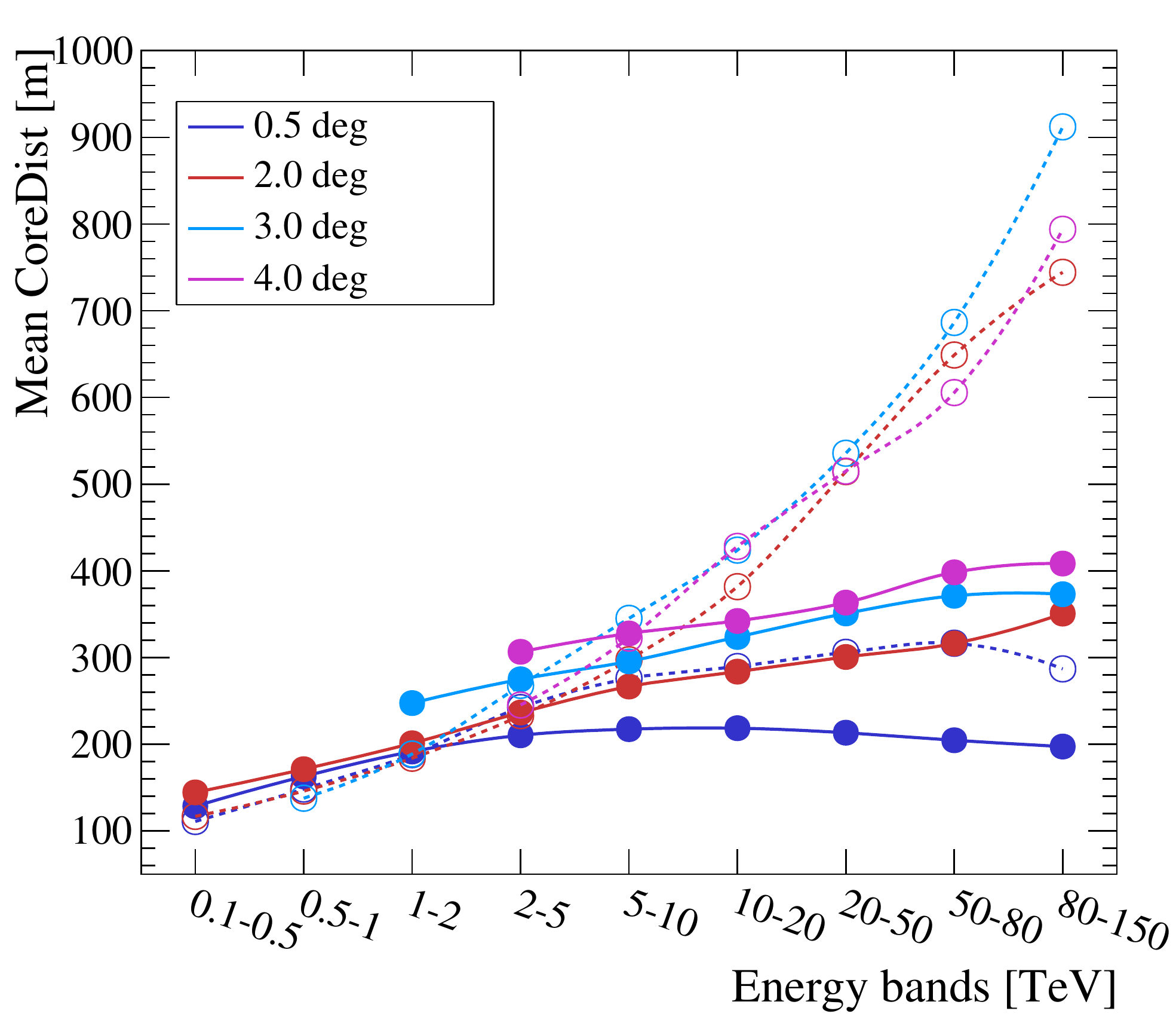}
  \end{subfigure}
    \begin{subfigure}[b]{0.48\linewidth}
    \includegraphics[trim=0 0 0 18, clip, width=1.0\linewidth]{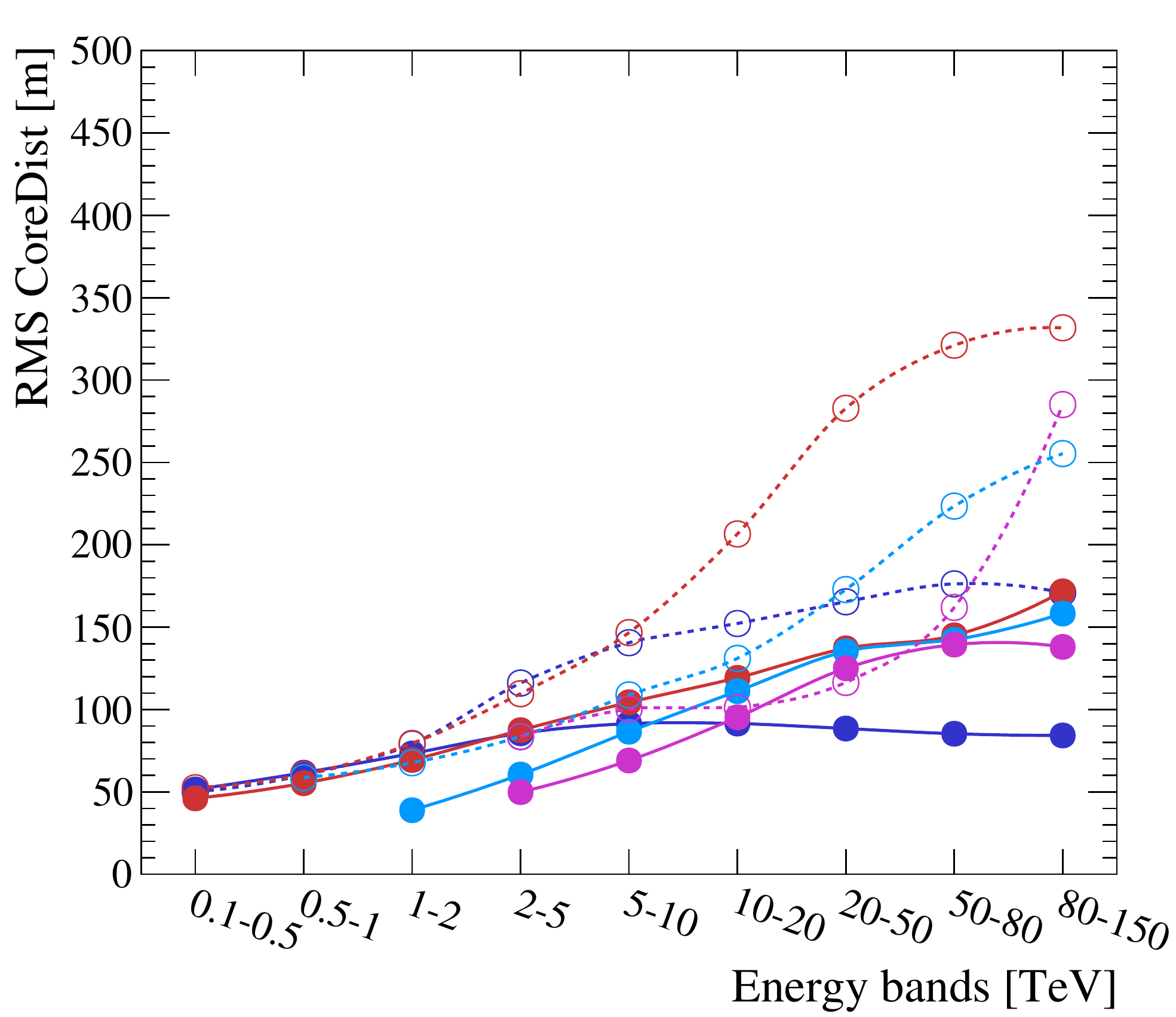}
  \end{subfigure}
  \caption{A comparison of the shower core distance parameter distribution behaviour for gamma-ray and background samples. The curves display the mean (left) and RMS (right) values of the core distance distributions as a function of energy. Solid lines denote point-like gamma rays simulated at 20$^{\circ}$ zenith, while dashed ones indicate real observations of the empty fields, i.e. cosmic-ray background, in the zenith angle range between 15$^{\circ}$ and 25$^{\circ}$.}
  \label{fig:fig4}
\end{figure}

The background rejection in this work is based on the standard method implemented within the H.E.S.S. analysis framework \cite{gh_ohm}. It uses a machine learning algorithm, specifically Boosted Decision Trees (BDT) implemented in the TMVA package. In addition to the standard list of six input variables that consists of four image-shape-based parameters, shower maximum and uncertainty of energy reconstruction, the method presented here uses one more input parameter, which showed background discrimination power. This parameter is the shower core distance, i.e. the distance between the shower axis and array centre. The comparison of the shower core distance distribution for gamma rays and background events is shown in Figure \ref{fig:fig4}. As can be seen in the figure, the core distance distributions are similar at low energies. However, at energies above $\sim$5\,TeV, hadrons have a broader distribution of core distances and a significant fraction of events are reconstructed to further distances in comparison to gamma rays.

Similarly to the standard method, the BDTs in this work are trained in energy, zenith and offset angle bands. The list of zenith angle bands stays unchanged and is the same as for the neural network training in the DISP method (see Table \ref{tab:disp_binning}). The energy and offset angle bands are modified. Instead of one offset-angle band at 0.5$^{\circ}$, there are nine bands, which cover the range from 0.5$^{\circ}$ to 4.5$^{\circ}$ with the step of 0.5$^{\circ}$. Instead of one band at high energies from 5 to 100\,TeV, five new energy bands are introduced: 5\,--\,10, 10\,--\,20, 20\,--\,50, 50\,--\,80, 80\,--\,150\,TeV.

Analogically to the standard method, the background sample for the training is extracted from the real observations of the sky regions, which do not contain bright gamma-ray sources, while point-like gamma-rays for the signal sample are simulated. The energy range covered in the simulation is the same as for the gamma rays used for the training in the DISP method. In order to have large gamma-ray statistics at high energies, the simulated spectra again have $\Gamma = -1$ spectral index instead of the standard $\Gamma = -2$. However, due to such a choice of the simulated spectra, there is not enough statistics at low energies and two low energy bins and combined into one 0.1\,--\,0.5\,TeV bin. This will affect the classification performance in the low energy band, but it is considered acceptable since the current work is focused on high energies. 

Eventually, the modifications introduced to the standard gamma-hadron separation method improved the background rejection efficiency above 10\,TeV at the lowest offset angles, improved effective area at intermediate offset angles (although, at the cost of slightly increased background rate) and allowed us to use large offset angles in the analysis as is shown in the next section.


\section{Performance}
\label{sec:sec3}

This contribution focuses on general method development, while the cut-optimisation investigation is beyond the scope of this work. Thereby, the analysis proposed in this study is based on the standard cut configuration and uses the same values for the event selection and gamma-hadron separation cuts. 

\begin{figure}[b]
  \centering
  \begin{subfigure}[b]{0.48\linewidth}
    \includegraphics[trim=0 0 0 20, clip, width=1.0\linewidth]{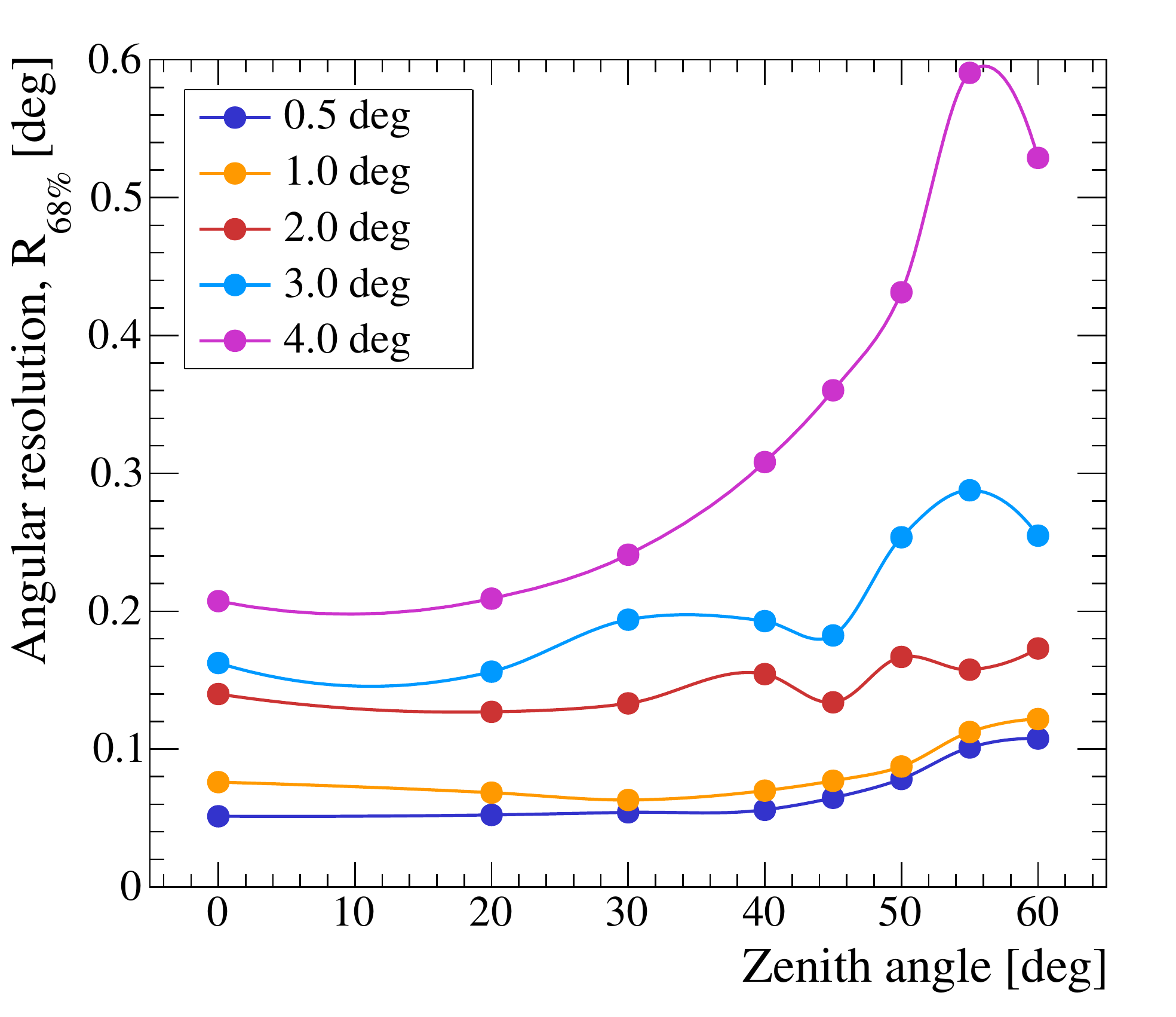}
  \end{subfigure}
    \begin{subfigure}[b]{0.48\linewidth}
    \includegraphics[trim=0 0 0 20, clip, width=1.0\linewidth]{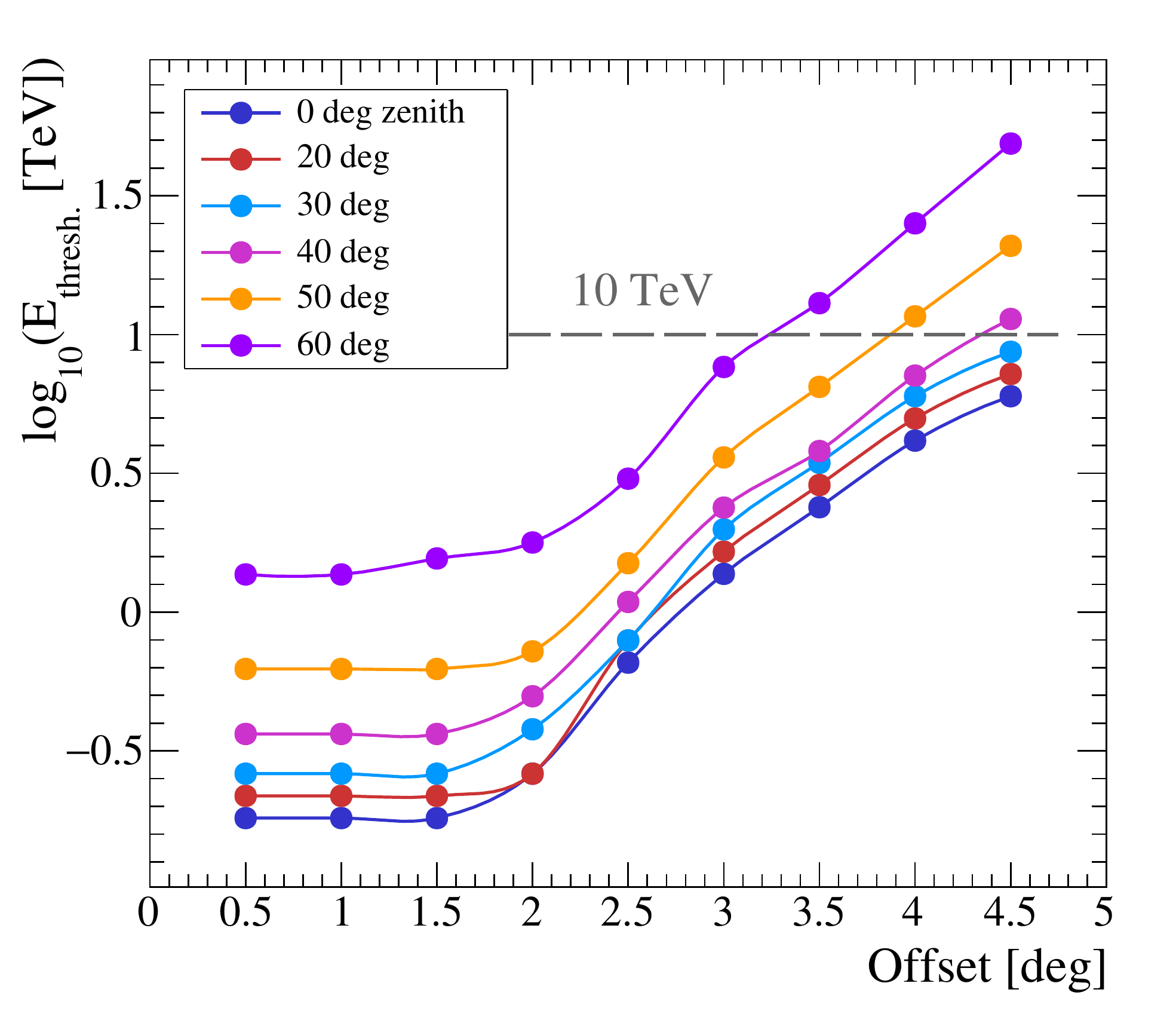}
  \end{subfigure}
  \caption{The angular resolution above 10\,TeV and energy threshold of the high-energy optimised analysis as a function of zenith and offset angle.}
  \label{fig:fig5}
\end{figure}

The left side of Figure \ref{fig:fig5} shows the angular resolution at different event offsets as a function of the zenith angle. Points represent the angular resolution estimated for energies above 10\,TeV. For zenith angles above 50$^{\circ}$, the energy threshold is higher than 10\,TeV for the offset angle of 4$^{\circ}$. In this case, the angular resolution is computed above the energy threshold, which also can be seen on the right side of Figure \ref{fig:fig5} as a function of offset and zenith angle. Within the physical size of the camera, the energy threshold stays nearly constant for a given zenith angle. But at offset angles that are larger than the camera size, the energy threshold experiences a dramatic increase.

\begin{figure}[t]
  \centering
  \begin{subfigure}[b]{0.48\linewidth}
    \includegraphics[trim=0 0 0 18, clip, width=1.0\linewidth]{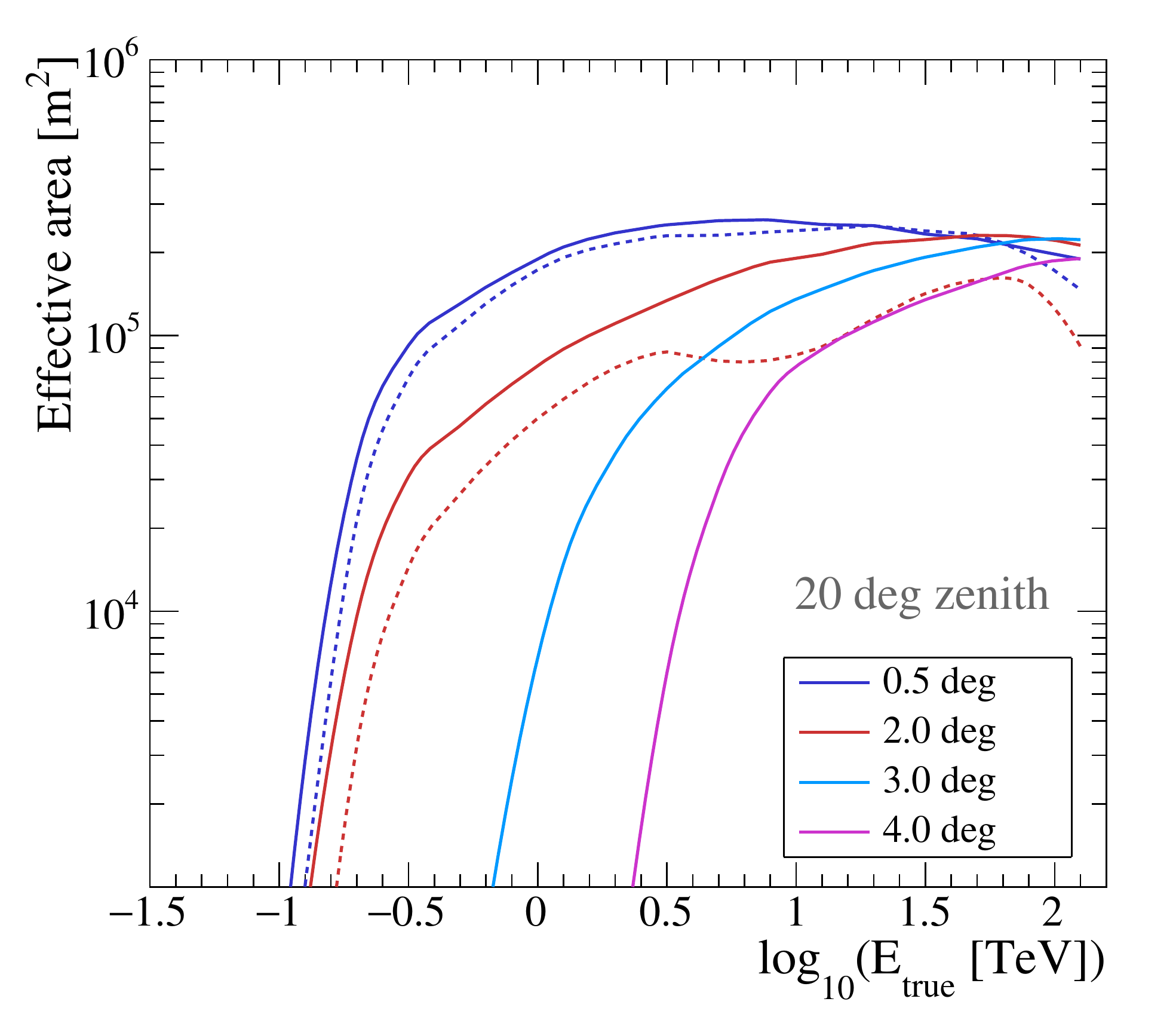}
  \end{subfigure}
    \begin{subfigure}[b]{0.48\linewidth}
    \includegraphics[trim=0 0 0 18, clip, width=1.0\linewidth]{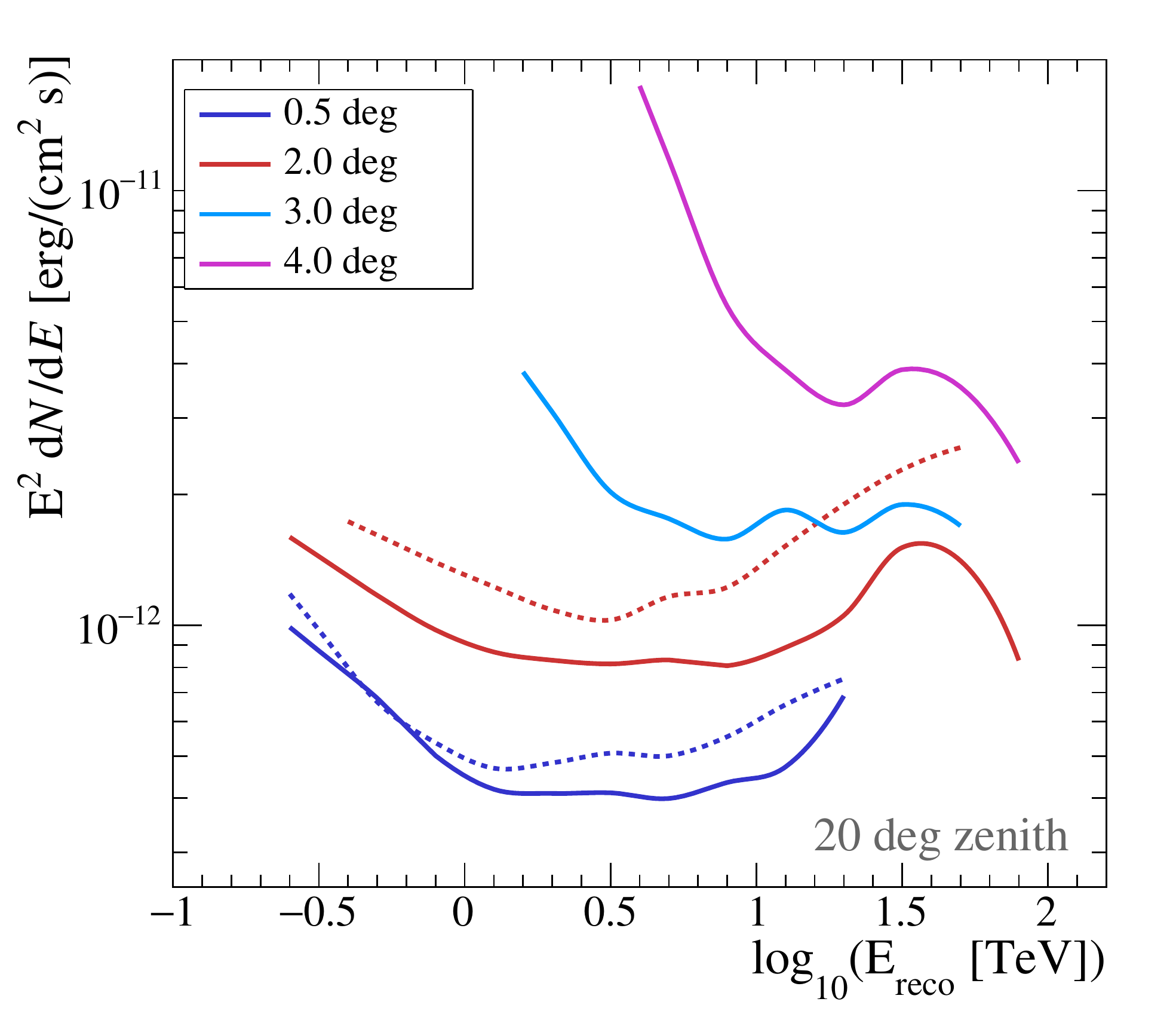}
  \end{subfigure}
  \caption{The effective area and differential sensitivity for the 100\,h of observations as a function of energy for 20$^{\circ}$ zenith angle and different offset angles. Solid and dashed lines denote the curves for the high-energy and the standard analysis, respectively. The x-axis is binned a way to contain 5 energy bins per decade.}
  \label{fig:fig8}
\end{figure}

Figure \ref{fig:fig8} shows the effective area in the left and differential sensitivity of the developed analysis method in the right panel and compares the results to the standard analysis for the events at small offset angles. At 0.5$^{\circ}$ offset angle, the effective area curves for the two analyses are very similar, which is expected since all selection cut values are the same. However, at larger offset angles, the standard effective area experiences a depression around 5\,--\,10\,TeV. It might indeed happen due to the fact that the standard gamma-hadron separation is trained only for the 0.5$^{\circ}$ offset band and performs worse at larger offset angles. The sensitivity curves in the figure show an improvement at the level of 10\,--\,20\% in comparison to the standard analysis. At the lowest offset angles, the sensitivity improvement is reached as a result of the more efficient background rejection, while at the intermediate offset angles, it is achieved thanks to the improved effective area.

This contribution also presents the performance studies carried out by applying the developed high-energy analysis to well known gamma-ray sources. The results are displayed in Figure \ref{fig:fig8}, which shows the evolution of the significance and gamma-ray excess as a function of the maximum allowed event offset (2$^{\circ}$, 3$^{\circ}$ and 4$^{\circ}$) in the high-energy optimised analysis for the four selected sources. In addition, the results are compared to the standard analysis with maximum event offset set to 2$^{\circ}$. On average, high-energy analysis with 3$^{\circ}$ and 4$^{\circ}$ maximum event offset has the highest values of excess counts, which is important for the  morphology and spectral studies. However, these analysis configurations also typically have a higher background rate after gamma-hadron separation, which results in high-energy analysis with 2$^{\circ}$ and 3$^{\circ}$ maximum event offset having higher significance as well as signal-to-noise values. When compared to the standard analysis, 2$^{\circ}$-high-energy analysis performs better. 


\section{Conclusion}
\label{sec:sec4}

In this contribution, an high-energy-optimised analysis method for IACTs is presented. It allows for the reconstruction of events at large-offset angles, which are typically not accessible in the standard analysis. The inclusion of the large-offset events in the analysis can increase the effective FoV up to 9$^{\circ}$, and more importantly, it is a promising way to increase the source exposure, and as a result, valuable gamma-ray statistics above 10\,TeV. 

The challenges of degraded reconstruction accuracy and increased background rate encountered during this study are addressed by implementing the DISP method for better event direction reconstruction as well as the introduction of several improvements to the standard gamma-hadron separation method. Altogether, presented analysis modifications allows not only for the usage of large-offset events but also resulted in a 10\,--\,20\% improvement in sensitivity at standard offset angles, i.e. below 2.5$^{\circ}$.


\begin{figure}[t]
  \centering
  \begin{subfigure}[b]{0.49\linewidth}
    \includegraphics[trim=0 0 0 20, clip, width=1.0\linewidth]{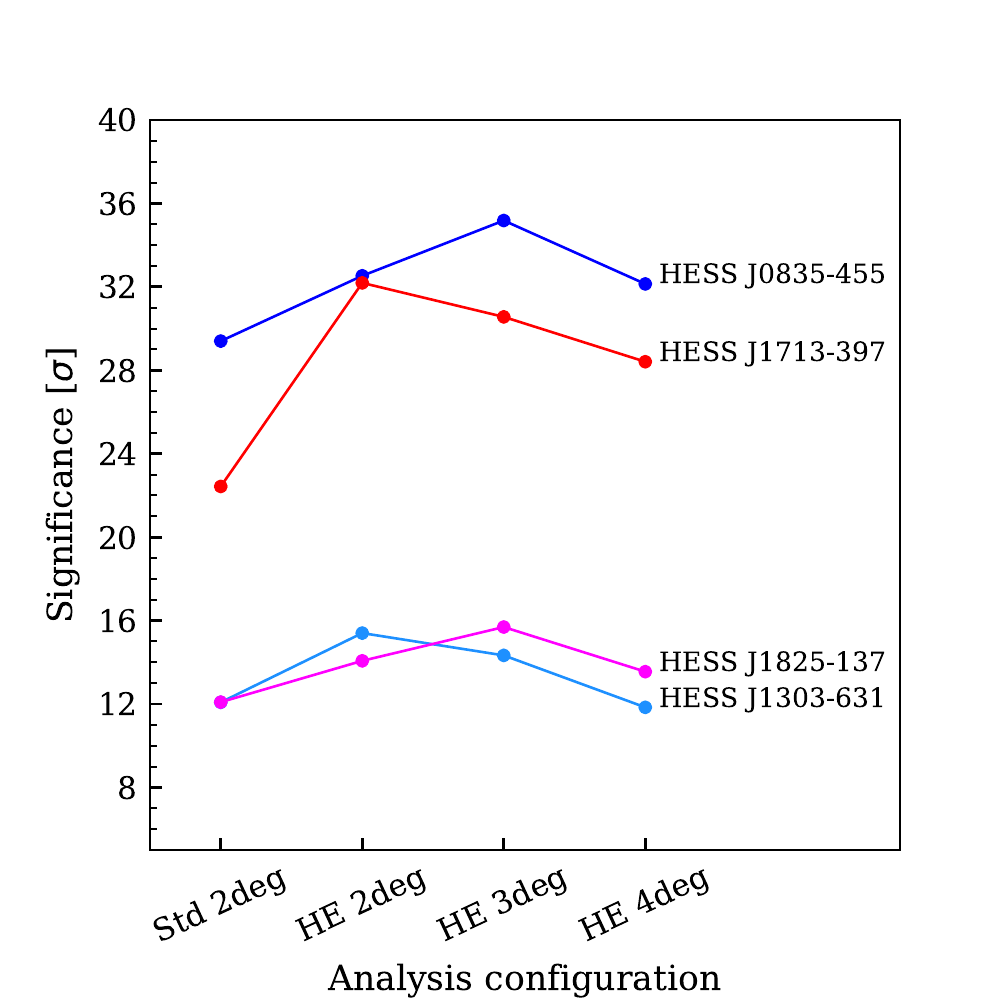}
  \end{subfigure}
    \begin{subfigure}[b]{0.49\linewidth}
    \includegraphics[trim=0 0 0 20, clip, width=1.0\linewidth]{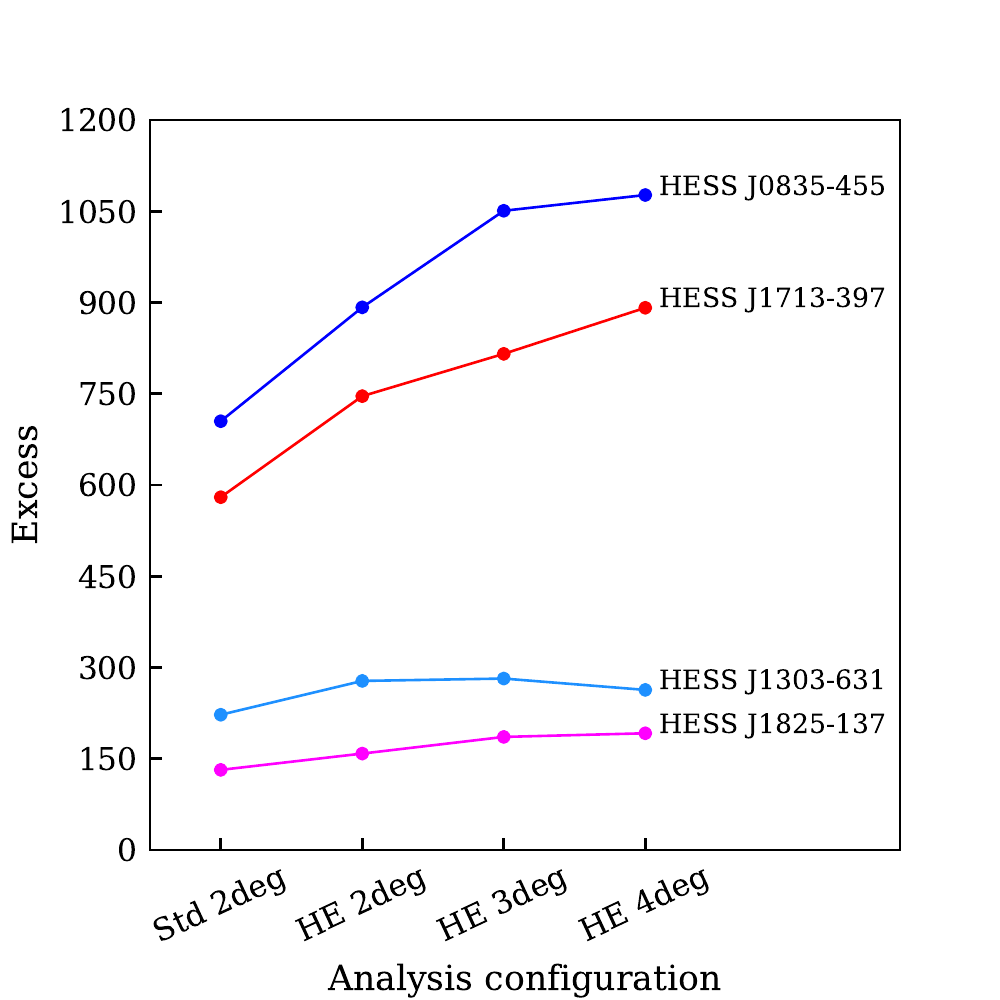}
  \end{subfigure}
  \caption{Significance (left) and excess counts (right) for different analysis configurations: high-energy analysis with three different maximum event offset and comparison to the standard analysis.}
  \label{fig:tev_cat_map}
\end{figure}

\section{Acknowledgments}
We thank the H.E.S.S. Collaboration for allowing us to use H.E.S.S. data and analysis software package for this contribution.

%
%
%

\end{document}